\documentclass[pra,twocolumn,amsfonts,amsmath,floatfix,superscriptaddress,nofootinbib]{revtex4-1}

\usepackage{graphicx}
\usepackage[usenames,dvipsnames,svgnames,table]{xcolor}
\usepackage[caption=false]{subfig}
\graphicspath{{./figures/}}
\usepackage{dcolumn}
\usepackage{bm}
\usepackage{hyperref}
\usepackage{multirow}

\usepackage{amsmath}

\makeatletter
\let\@@pmod\pmod
\DeclareRobustCommand{\pmod}{\@ifstar\@pmods\@@pmod}
\def\@pmods#1{\mkern4mu({\operator@font mod}\mkern 6mu#1)}
\makeatother

\newcommand{\Bra}[2][]{\left<#2\right|_{#1}}
\newcommand{\Ket}[2][]{\left|#2\right>_{\hspace{-0.1em}#1}}

\newcommand{\Ketbra}[3][]{\left|#2\middle>\middle<#3\right|_{#1}}

\newcommand{\Proj}[2][]{\Ketbra[#1]{#2}{#2}}

\newcommand{\ba}[1]{\begin{array}{#1}}
\newcommand{\ea}{\end{array}}

\newcommand{\lr}[1]{\left( #1 \right)}
\newcommand{\ddd}{\mathrm{d}}

\begin{document}
\preprint{APS/123-QED}

\title{High-dimensional quantum encoding via photon-subtracted squeezed states}

\date{\today}

\begin{abstract}

We introduce a high-dimensional quantum encoding based on coherent mode-dependent single-photon subtraction from multimode squeezed states. This encoding can be seen as a generalization to the case of non-zero squeezing of the standard single-photon multi-rail encoding. The advantage is that the presence of squeezing enables the use of common tools in continuous-variable quantum processing, which in turn allows to show that arbitrary $d$-level quantum states can be generated and detected via simply tuning the classical fields that gates the photon-subtraction scheme. Therefore, the scheme is suitable for mapping arbitrary classical data in quantum mechanical form.  Regardless the dimension of the data set alphabet, the mapping is conditioned on the subtraction of a single photon only, making it nearly unconditional. We prove that this encoding can be used to calculate vector distances, a pivotal primitive in various quantum machine learning algorithms.

\end{abstract}

\author{Francesco Arzani}
\email{fra.arzani@gmail.com}
\affiliation{Universit\'e de Lorraine, CNRS, Inria, LORIA, F 54000 Nancy,
France}
\affiliation{LIP6, CNRS, Sorbonne Universit\'e, 4 place Jussieu, 75005 Paris, France }

\author{Alessandro Ferraro}
\email{a.ferraro@qub.ac.uk}
\affiliation{Centre for Theoretical Atomic, Molecular and Optical Physics, Queen’s University Belfast, Belfast BT7 1NN, United Kingdom; }

\author{Valentina Parigi}
\affiliation{Laboratoire Kastler Brossel, Sorbonne Universit\'e, CNRS, ENS-PSL Research University, Coll\`ege de France; CC74, 4 Place Jussieu, 75252 Paris, France;}

\date{\today}

\maketitle

\section{Introduction}

A crucial aim of the research on quantum information technologies is to harness quantum systems so that some information processing tasks can be achieved with better performances than it is possible with classical computers~\cite{NChuang}. To exploit the improvement predicted by the theory it is necessary to encode information on a physical system whose quantum properties can be preserved and controlled in the laboratory. With this aim, much effort has been devoted to investigate the use of light as  a carrier of quantum information, due to both its robustness to noise and the availability of advanced technological tools to control its state (classical or quantum) \cite{Kok2007, Pan2012, Ansari18}. 

As first step in any information processing task acting on a classical input --- be it communication, data processing, or universal computation --- an encoding must be chosen to \textit{write} the input information on the given system. To this end, a correspondence must be established between the possible inputs and a subset of states of the carrier system. The choice of the encoding then determines the physical implementation that corresponds to the \textit{logical} processing of information. Light is described in quantum mechanics by an infinite-dimensional Hilbert space~\cite{leonhardt1997measuring, walls2007quantum}, whereas common communcation or computational tasks are defined in terms of finite alphabets~\cite{NChuang}. A common choice is then to encode information in a finite-dimensional subspace of a single mode, such as that spanned by states with a finite number of photons (Fock states) or with a definite single-photon property (\textit{e.g.} polarization or orbital angular momentum states). A much celebrated variation to these schemes involves multiple modes instead, and it is given by the so-called dual-~\cite{multirail1} or multi-rail~\cite{Kok2007} encoding. In the latter, information is stored in the presence or absence of photons in each of a set of spatial or temporal modes of the radiation. 
All these encodings belong to the realm of what is commonly known as discrete-variable (DV) quantum information (finite-dimensional quantum systems). 
However, states with a definite number of photons can, as of now, be produced only probabilistically, making the setup not easily scalable. Moreover, protocols devised within this paradigm often require photon counting, which is experimentally demanding, especially when high efficiency is required.

An alternative approach is to encode information using the whole infinite-dimensional Hilbert space. The typical observables of interest are then the quadratures of the field, akin to mechanical position and momentum, which have a continuous spectrum. For this reason, this choice corresponds to the so called continuous-variable (CV) regime~\cite{Braunstein:05, cerf2007quantum, serafini2017quantum}. Among the advantages of the latter are the facts that \textit{(i)} entangled and non-classical states can be produced deterministically using squeezed states and \textit{(ii)} states can be detected using the highly efficient scheme of homodyne measurements \cite{Ferraro05, Weedbrook2011}. On the other hand, the mathematics becomes considerably more involved due to the need of dealing with infinite dimensions, and the direct correspondence with finite-dimensional logical qubits is lost. 

A common way to recover such correspondence is to encode a single logical qubit into a single infinite-dimensional system by using the symmetries of certain states --- for example, the translational symmetry of GKP states  (introduced by Gottesman, Kitaev and Preskill) \cite{Gottesman2001} or the parity symmetry of cat and binomial states \cite{Chuang1997, Lund2008, Michael2016}. Such strategies allow to use a discrete alphabet that nevertheless can be manipulated using the mathematical and experimental machinery typically applied in the CV approach. In this work, we move from such strategies and, rather than encoding single DV systems in single CV systems, we propose to encode the former into multiple instances of the latter. In this sense, we introduce a CV counterpart of the standard DV dual- and multi-rail encoding. More formally, as it will be described below, the zero squeezing limit of our encoding corresponds to the DV multi-rail encoding. 

Here, we will focus on states that are produced by coherently\footnote{Here the word \textit{coherently} refers to the fact that the single photon subtraction is applied on a coherent superposition of modes. This is not the same meaning as in ``coherent states'', also common in quantum optics. In the following, the distinction should be clear from the context. \label{coh} }  subtracting one photon from several modes that are each in a squeezed state. Our interest is mainly motivated by recent theoretical~\cite{Averchenko14,Averchenko2016} and experimental~\cite{Eckstein11,Brecht14,Rosenblum2015,Ra17,jacquard,inprep} advances in the production of this kind of states, that showed how to coherently subtract single photons from multi-mode squeezed states via the interaction in a non-linear crystal with an appropriate classical field. 

We develop our analysis along two main directions. First, we study how quantum information can be encoded in the multi-mode code space corresponding to a single qudit ($d$-dimensional quantum system) or an ensemble of qubits. The advantage of our scheme with respect to the usual multi-rail DV approach lies in the fact that the presence of squeezing enables the use of common tools in continuous-variable quantum optical processing. This, in turn, allows us to show that arbitrary qudits can be generated and detected via simply tuning the classical fields that drive the photon-subtraction scheme. We also investigate how parity measurements --- which can be related to homodyne measurements --- can be used to discriminate between basis states in the multi-mode code space. 

The resilience of the proposed scheme with respect to the main noise mechanisms is also analyzed, finding that high levels of squeezing make the code space less resilient to losses. 

The second direction we explore consists in considering the mapping of classical strings of data on the photon-subtracted state. In general, mapping classical data into quantum states constitutes an unavoidable initial step, which is essential to any further quantum processing of the input, including the proper evaluation of the processing performances \cite{Aaronson2015, Zhao2018c}. In our case, this mapping is enabled by the fact that, as said, arbitrary high-dimensional states can be generated by tuning the classical gate fields which, in turn, can directly encode the classical data. Remarkably, the fact that only one probabilistic event is needed to produce a logical state (regardless the size of the input data alphabet) implies that the classical to quantum mapping is nearly unconditional. We propose two protocols that exploit this mapping to compute either the scalar product or the distance between classical data vectors by measuring a single quadrature of the field. The interest here lies in the wide-spread application of these primitives for quantum-enhanced machine learning schemes \cite{Schuld2015, Biamonte2016a, Dunjko2017}, \textit{e.g.} in distance-based clustering algorithms for supervised pattern recognition \cite{Schuld2017c}. 

The rest of the article is structured as follows. In section~\ref{sec:photoSub} we recall the physics of mode-dependent single photon subtraction. We then describe the encoding in section~\ref{sec:encoding}, where we also detail how information can be extracted by measurements that discriminate between elements of the computational basis, and how the main sources of noise affect the encoding.  Section~\ref{sec:classStrings} is devoted to the encoding of classical data and the computation of scalar products and vector distances. Conclusive remarks in section~\ref{sec:end} complete the paper. Appendix~\ref{App:states} reports some examples of interesting encoded states while in  Appendix~\ref{App:gates} a universal set of operations is defined that could be used to process the encoded information.

\section{Mode dependent single-photon subtraction \label{sec:photoSub}}

We now introduce some notations and recall how coherent photon subtraction from squeezed time-frequency modes works. Consider a multi-mode squeezed state (MMSS) $\Ket{S}$, composed of $M$ modes and written as 
\begin{equation}
\Ket{S} = \bigotimes_{j=1}^{M} \Ket[e_j]{s_j} \;,\label{eq:MMSS}
\end{equation}
where each $\Ket[e_j]{s_j}=S(s_j)\Ket[e_j]{0}$ is a squeezed state of mode $e_j\lr{\boldsymbol{x},\omega}$ with squeezing parameter $s_j$, $\lbrace e_j\lr{\boldsymbol{x},\omega}\rbrace$ being an orthonormal set of modes of the electric field, which are functions of the position $\boldsymbol{x}$ and frequency $\omega$. We denote by $\sigma_j$ the corresponding annihilation operators. States in Eq.~(\ref{eq:MMSS}) have been experimentally realized in various contexts \cite{suClus, armstrong12, Roslund14, Chen14, Yokoyama13}, producing and detecting quantum states across up to a million modes \cite{million}. As a relevant example, we bear in mind the case of MMSS generated by parametric down-conversion of a frequency comb. In particular, in the simple case of a comb with a Gaussian spectrum, the spectra of the co-propagating squeezed modes can be approximated with Hermite-Gauss functions \cite{Roslund14,patera}. A sum-frequency conversion process can then be used to up-convert part of the light from a mode $f$ defined as
\begin{equation}
f\lr{\boldsymbol{x},\omega} = \sum _j c_j^* e _j\lr{\boldsymbol{x},\omega}
\end{equation} 
with $c_j\in\mathbb{C}$, $\sum _j |c_j|^2 =1$. This can be accomplished by mixing the MMSS with a strong coherent pulse (gate field) in a non-linear crystal \cite{Averchenko14}. By choosing the phase matching conditions for a non-collinear configuration, the up-converted light is emitted in a different direction with respect to the transmitted MMSS and gate beams. Thus, the process can be modeled as an effective weak beam splitter interaction \cite{Averchenko2016}. The activation of a single-photon detector [\textit{e.g.}, an avalanche photodiode (APD)] placed on the path of the up-converted signal can then herald the subtraction of a photon from the MMSS. If the process is perfect, the photon comes with certainty from mode $f$.  As it was theoretically shown in \cite{Averchenko14} and recently experimentally demonstrated \cite{Ra17}, with the appropriate phase-matching conditions and disergarding the spatial dependence, the spectrum of the mode $f$ essentially coincides with the spectrum of the gate field. The corresponding annihilation operator is thus defined by
\begin{equation}
b = \sum _j c_j \sigma _j\;, \label{eq:subMode}
\end{equation} 
and the state of the transmitted MMSS after the detection of a photon is a multimode photon-subtracted (MMPS) state:
\begin{equation}
\Ket{MMPS}  = \sum_j \gamma_j \Ket[e_j]{s_j ^p} \bigotimes _{i\neq j } \Ket[e_i]{s_i}\;.\label{eq:psub_MMSS}
\end{equation} 
Here $  \Ket[e_j]{s_j ^p} = \mathcal{N}_j \sigma_j \Ket[e_j]{s_j }$ denotes a photon-subtracted state with $\mathcal{N}_j$ a normalization factor. Due to the factors $\mathcal{N}_j$, the complex coefficients $\gamma_j$ do not coincide with the $c_j$ ones, however they are fully determined by the latter. When the normalization factors $\mathcal{N}_j$ are taken into account, it is then possible to find the appropriate  gate that will produce a photon-subtracted state as in Eq.~(\ref{eq:psub_MMSS}) with arbitrarily chosen coefficients $\gamma_j$. 

\section{Encoding \label{sec:encoding}}

The photon-subtracted states in Eq.~(\ref{eq:psub_MMSS}) can be re-written as
\begin{equation}
\Ket{MMPS} \equiv \sum_{j=1}^{M} \gamma_j \Ket{j}\;, \label{eq:pureQudit}
\end{equation} 
where $\Ket{j}$ represents a MMSS in which a single photon has been subtracted from mode $e_j$:
\begin{equation}
\Ket{j}  = \Ket[e_j]{s_j ^p} \bigotimes _{i\neq j } \Ket[e_i]{s_i} \;.
\label{eq:psub_basis}
\end{equation} 
The relevant observation here is that the states $\Ket{j}$ are orthogonal since they belong to subspaces with definite parity, in terms of photon population. In particular, one has that \begin{equation}
\langle i | j \rangle \propto   \Bra[e_i]{s_i}  \Ket[e_i]{s_i^p}   \Bra[e_j]{s_j ^p}  \Ket[e_j]{s_j} = 0 \;,
\end{equation} since squeezed states only contain even photon components and photon-subtracted squeezed states only contain odd photon components. We will refer to states $\Ket{j}$ as the \textit{computational basis}, each value $j$ corresponds to a symbol of a finite alphabet, and Eq.~(\ref{eq:pureQudit}) introduces an abstract notation for an encoded qudit. If the number of modes in $M$ is $2^n$, the
$ \mathrm{span} \left\{\Ket{j}\right\} $ 
is isomorphic to $\mathbb{C}^{2n}$ and the qudit can be thought to represent $n$ qubits. 

Experimental tomography of the photon-subtraction process demonstrated a purity of more than $90\%$ for the superposition of 16 modes at different frequencies~\cite{Ra17}. Hence, it is in principle possible to realize a highly accurate single-mode subtraction with a large experimental tunability of the coefficients $\gamma_j$ in Eq.~(\ref{eq:pureQudit}) which, in particular, is nearly independent of the number of modes in the system. This means that any, ideally arbitrary, state of a qudit with dimension $M$ can be generated. Equivalently, any superposition of $n$ qubits can be realized with a single photon subtraction from $M=2^n$ modes. Some examples can be found in Appendix~\ref{App:states}. Clearly, the number of modes scales exponentially with the number of qubits, but the number of single-photon operations needed is constant, namely equal to one. 

As mentioned, the present encoding can be regarded as a generalization of the usual DV multi-rail encoding. There, a single photon is prepared in an arbitrary superposition of $M$ spatially separated modes via a passive interferometer. The relation to the encoding here introduced stems from the fact that a photon-subtracted squeezed state is equivalent to a squeezed single photon states, namely:
\begin{equation}
\Ket[e_j]{s_j ^p} = \mathcal{N}_j \sigma_j S(s_j)\Ket[e_j]{0}\equiv\mathcal{N}_j' S(s_j)\Ket[e_j]{1}\;.
\end{equation}
As a consequence, the encoded state $\Ket{MMPS}$ in Eq.~\eqref{eq:psub_MMSS} represents an arbitrary single-photon superposition over $M$ modes (that could in principle be spatially separated) to which a multiple squeezing operator $\otimes_{j=1}^M S(s_j)$ has been applied. In the zero-squeezing limit, $\Ket{MMPS}$ states thus correspond in fact to the usual multi-rail encoding\footnote{Notice that in practice the scheme here introduced cannot be used to generate multi-rail encoding tough, since the probability of subtracting a photon vanishes for zero squeezing. \label{limit} }. 

We recall that, in standard multi-rail encoding, the interferometer parameters need to be set accordingly to the state to be generated, a procedure that is typically hard to implement especially in bulk optics or whenever a high degree of tunability is required, and it  can be implemented only via advanced integrated devices \cite{Carolan711,Xia18,Harris17}. The scheme presented here overcomes these issues entirely, since the superposition determining the MMPS states can be set simply by tuning the gate field parameters ---namely, modifying the spectral components of the strong coherent gate that drives the non-linear crystal \cite{Averchenko14}. In other words, due to the experimental possibility to select the subtracted mode, many resource states could be generated without modifying the physical setup. The limitation in the code dimension is the number of modes which can be simultaneously squeezed and addressed  by photon subtraction  \cite{Roslund14,Cai17,Ra17,jacquard}. The code rate is given by the succes rate of the single-photon subtraction: it is mainly driven by the gate power which has been limited to $\sim 2$ KHz \cite{Ra17} in order to avoid spurious or multi-photon events. 

 Another interesting feature of this scheme is that single photon detectors are not necessarily needed, given that all modes are populated. In fact, as we will show below, highly-efficient homodyne detection can be used at the measurement stage. In addition, the detected modes can be tuned by appropriately selecting the local-oscillator fields for homodyning, as shown experimentally in Ref.\cite{Roslund14,Cai17}. This in turn implies that a variety of different unitary transformations (in particular, any mode mixing operation \cite{Cai17,Fer16}) can be implemented on the generated states by embedding them in the measurement.

\subsection{Distinguishing elements of the computational basis}

We saw that the parity features of the computational basis elements imply that the latter form an orthonormal set of states. In principle, it is thus possible to perfectly discriminate them. However, experimental imperfections will lead to a partial overlap and consequently a non-zero probability of failing to distinguish these states. The two most relevant kinds of imperfections in this context are non-ideal photon-subtractions and losses before the detection. The former is due to the incomplete mode-selectivity of the subtraction process, and it has been already  considered for the characterization of the experimental platform in Ref.~\cite{Ra17}. This is also the easiest to describe, because it only shuffles modes within the finite-dimensional code-space spanned by the basis elements. Losses, on the other hand, need to be described in the infinite-dimensional Hilbert space of the EM field. We shall analyze the two types of errors separately in the following.

Many figures of merit can be used to assess the performance of state discrimination. We will focus on the state fidelity under a specific kind of measurement based on the parity operator (defined below). The latter is a sensible choice due to the parity properties of the computational-basis state. 
In particular, we define 
\begin{equation}
J = \sum \limits_j \lr{ j{\tilde{\Pi}_j} }\label{eq:J} \;,
\end{equation} 
with 
\begin{equation}\label{eq:parity_tilde}
\tilde{\Pi}_j = \sum_k \Proj[j]{2k+1}\otimes \mathbb{I}_{\bar{j}}
\end{equation} 
a modified parity operator on mode $j$, that acts as the identity on all other modes. Note that $\tilde{\Pi}_j = \tilde{\Pi}_j^\dagger = \lr{\tilde{\Pi}_j}^2$, so $\tilde{\Pi}_j$ is a projector. Therefore, we have \begin{equation}
J\Ket{j} = j \Ket{j}.
\end{equation}

Even if it is experimentally challenging to measure the parity of many modes, the mean value of the parity operator \begin{equation}
\Pi_j = \sum_n (-1)^n \vert n \rangle \langle n \vert_j \otimes \mathbb{I}_{\bar{j}}  = 1 - 2 \tilde{\Pi}_j
\end{equation} is related to the value of the Wigner function at the origin of the phase space and can be measured through homodyne detection. In fact, given a generic single-mode state $\rho$, the Wigner function evaluated at point $\alpha$ of the phase space is given by \cite{Ban99,Royer77} (since we are dealing with a single mode, we drop the subscript $j$ for clarity)
\begin{equation} 
\begin{aligned}
W(\alpha)& = \frac{2}{\pi} \langle  D(\alpha) \Pi D(\alpha)^\dagger\rangle \\
& = \mathrm{Tr}[\rho \;\frac{2}{\pi}\sum_n (-1)^n D(\alpha)\vert n \rangle \langle n \vert D(\alpha)^\dagger] \\
 & = \frac{2}{\pi}\sum_n (-1)^n p_n(\alpha)\;, 
 \end{aligned}
\end{equation} 
where $p_n(\alpha)$ is the occupation probability of the $n$-photon state after the state $\rho$ has undergone a phase-space displacement operation $D(-\alpha)$ by an amount $-\alpha$ \footnote{Recall that $\langle  D(\alpha) \Pi D(\alpha)^\dagger\rangle = \mathrm{Tr}[\rho  D(\alpha) \Pi D(\alpha)^\dagger ] = \mathrm{Tr}[ D(\alpha)^\dagger \; \rho \;  D(\alpha) \Pi] = \mathrm{Tr}[ D(-\alpha) \; \rho \;  D(-\alpha)^\dagger \Pi]$}. The extension to the multi-mode case is trivial. The average value of the parity operator, which is sufficient to discriminate the basis elements $\Ket{j}$, coincides with no displacement as 
\begin{equation} \begin{aligned}
W(0) &=\frac{2}{\pi} \langle \Pi\rangle\;. \end{aligned}
\end{equation}  
Direct evaluation of the Wigner function at any point in the phase space can be made by photon-counting after a displacement operation  \cite{Ban99,lai10},  but it can also be recovered more conveniently by cascaded optical homodyne, as proposed in  \cite{cascadedhomo}. In the particular case of the parity operator, where only the Wigner function at the origin of the phase-space is needed, the probability $p_n(0)$ can be simply obtained from phase-randomized homodyne measurements \cite{Mun95}. This is straightforward to realize experimentally and crucially the complexity of this measurement set-up (including the total number of measurements) increases only linearly with the number of modes $M$.

\subsubsection{Errors from imperfect photon subtraction \label{sec:fidePhotoSub}}

A perfect single-mode photon subtraction from one of the squeezed modes can be represented as a map \begin{equation}
\mathcal{P}_j : \Ket{S} \mapsto \mathcal{N}_j \sigma_j \Ket{S}\Bra{S}  \sigma_j ^\dagger = \Ket{j}\Bra{j}.
\end{equation} An imperfect subtraction from the mode $j$, that we denote by $\mathcal{I}_j$, can be modeled as a multimode process characterized by a subtraction matrix $\chi^{\lr{j}}$ such that~\cite{Ra17} \begin{equation}
\mathcal{I}_j: \Ket{S} \mapsto \sum_{kl} \chi^{\lr{j}}_{kl} \sqrt{\mathcal{N}_k\mathcal{N}_l}\sigma_k \Ket{S}\Bra{S}  \sigma_l ^\dagger = \sum_{kl} \chi_{kl} ^{\lr{j}}\Ket{k}\Bra{l}.
\end{equation} The matrix $\chi^{\lr{j}}$ has the same properties of a density matrix. In fact, it can be identified with the density matrix in the qudit basis of the state $\tilde{\rho}_j$, corresponding to a distorted element of the computational basis. The matrices $\chi^{\lr{j}}$ can be obtained experimentally via a tomography of the subtraction process. It is then easy to compute the fidelity between the ideal state $\Ket{j}$ and its realistic realization $\tilde{\rho}_j$ as \begin{equation}
\tilde{\mathcal{F}}_j\lr{j} = \Bra{j} \tilde{\rho}_j \Ket{j} = \chi^{\lr{j}}_{jj}\;,
\end{equation} or between $\tilde{\rho}_j$ and any basis state $\Ket{k} \neq \Ket{j} $ \begin{equation}
\tilde{\mathcal{F}}_k\lr{j} = \Bra{k} \tilde{\rho}_j \Ket{k} =  \chi^{\lr{j}}_{kk}.
\end{equation} This has the operational meaning of probability that $\tilde{\rho}_j$ will pass a test to check whether $\tilde{\rho}_j =  \Ket{k}\Bra{k} $, minimized over all possible measurement strategies ~\cite{wilde}. Note that $\Bra{k} \tilde{\rho}_j \Ket{k}$ is also the probability of getting the outcome $k$ when measuring $J$ on the state $\tilde{\rho}_j$, so $J$ optimally discriminates between computational-basis states, which \textit{a posteriori} justifies its definition. 

We can also easily compute the probability to get a wrong outcome $m\neq j$ when performing a measurement in the computational basis, as defined by $J$ in Eq.~(\ref{eq:J}), on a an imperfect basis state $\tilde{\rho}_j$: \begin{equation}
\sum_{m\neq j}\mathrm{Pr}\lr{J=m | \tilde{\rho}_j} = 1 - \mathrm{Tr}\lr{\tilde{\Pi}_j \tilde{\rho}_j} = 1 - \chi^{\lr{j}}_{jj}.
\end{equation} 
Note that the fidelity and the error probability do not depend on the squeezing level since, as mentioned above, the imperfection of photon subtraction considered here only shuffles states in the qudit-space, whose definition is independent of the amount of the squeezing in each mode.

We can then assess the robustness of computational-basis states  in Eq.~(\ref{eq:psub_basis}) to imperfect subtraction from the matrix $\chi^{\lr{l}}_{jk}$, which can be in turn determined experimentally via tomography of the subtraction process~\cite{Ra17, YSPC}. As an example, in the follwing we have considered data from Ref.~\cite{YSPC} (partially published in ~\cite{Ra17}). The values of $\chi^{\lr{j}}_{kk}$ measured from the tomography of the photon-subtraction from $M=4$ squeezed modes are reported in the following table:

\begin{center}
\begin{tabular}{cc|cccc|}
\multicolumn{2}{c}{} & \multicolumn{4}{c}{$k$} \tabularnewline
\cline{3-6}
 \multirow{6}{*}{\rotatebox{90}{$j$}} 
 & & 1  & 2 & 3  & 4 \\
\cline{2-6} 
&  \multicolumn{1}{|r|}{1} & .972 & .023 & .001 & .001 \\
& \multicolumn{1}{|r|}{2} & .031 & .932 & .032 & .002 \\
& \multicolumn{1}{|r|}{3}  & .007 &.045  & .893 & .046\\
& \multicolumn{1}{|r|}{4}  & .004 & .005  & .07 & .857 \\
\cline{2-6}
\end{tabular} \end{center} This shows that the error probability is below $15\%$ in the worst case.

The subsequent table shows instead the fidelity between two non-ideal basis states 
\begin{equation}
\tilde{\mathcal{F}}\lr{j,k} = \mathrm{Tr} \left[ \sqrt{ \sqrt{\tilde{\rho}_j} \tilde{\rho}_k \sqrt{\tilde{\rho}_j}}\right] ^2 
= \mathrm{Tr} \left[ \sqrt{ \sqrt{\chi^{\lr{j}}} \chi^{\lr{k}} \sqrt{\chi^{\lr{j}}}}\right] ^2 
\end{equation} 
which can also be taken as a measure of distinguishability, although the interpretation as error probability no longer holds for two generally mixed states. One obtaines:
\begin{center}
\begin{tabular}{cc|ccc|}
\multicolumn{2}{c}{} & \multicolumn{3}{c}{$k$} \tabularnewline
\cline{3-5}
 \multirow{5}{*}{\rotatebox{90}{$j$}} 
  &   & 2 & 3  & 4 \\
\cline{2-5} 
&  \multicolumn{1}{|r|}{1} & 0.1 & 0.01 & 0.01 \\
& \multicolumn{1}{|r|}{2} &  & 0.14 & 0.01 \\
& \multicolumn{1}{|r|}{3}  & &  & 0.19 \\
\cline{2-5} 
\end{tabular} \end{center} where again values closer to zero correspond to better distinguishability.

\subsubsection{Errors from losses before the detection}

Losses before the detection stage can occur due to actual optical losses, imperfect mode-matching or to finite quantum efficiency of the detectors. A simple model that is commonly used for these situations consists in assuming that each mode of the signal field is coupled through a beam splitter of transmittivity $\tau_j$ to an ancillary mode in the vacuum state, which is then traced out. For the sake of clarity, we make the simplifying assumption that losses affect each mode in the same way $\tau_j = \tau$.  The loss super-operator $\mathcal{L}_\tau$ acting on the multi-mode density matrix is then factorized as $\mathcal{L}_\tau=\mathcal{L}_{1,\tau}^{\otimes n}$ with each $=\mathcal{L}_{1,\tau}$ a single-mode loss operator with parameter $\tau$.
We denote by 
\begin{equation}
\bar{\rho}_j = \mathcal{L}_\tau \left[ \left| j \rangle \langle j \right| \right]
\end{equation} 
the $j$-th state of the computational basis after losses have occurred. Since computational-basis states are factorized in the basis of squeezed modes \begin{equation}
\left| j \rangle \langle j \right| = \left| s_1 \rangle _{e_1}\langle s_1 \right|  \otimes \ldots \otimes \mathcal{N}_j \left| s^p_j \rangle _{e_j}\langle  s^p_j \right|  \otimes \ldots \;,
\end{equation} 
the fidelity 
\begin{equation} \label{eq:FjjLoss}
\bar{\mathcal{F}}_j = \Bra{j} \bar{\rho}_j \Ket{j} 
\end{equation} 
between $\Ket{j}$ and $\bar{\rho}_j$ also factorizes as the product of fidelities of single-mode states: \begin{equation}
\begin{aligned}
\bar{  \mathcal{F }}_j =& \mathcal{N}_j^2  \Bra{j} a_j ^\dagger  \mathcal{L}_{1,\tau} \lr{    a_j \left| s_j \rangle \langle s_j \right|  a_j  ^\dagger}  a_j \Ket{j} \times \\& \prod_{k\neq j}   \Bra{s_k} \mathcal{L}_{1,\tau} \lr{  \left| s_k \rangle \langle s_k \right| }  \Ket{s_k}. \end{aligned}
\end{equation} 

\begin{figure*}

\begin{minipage}{0.92\textwidth}
        \includegraphics[width=\textwidth]{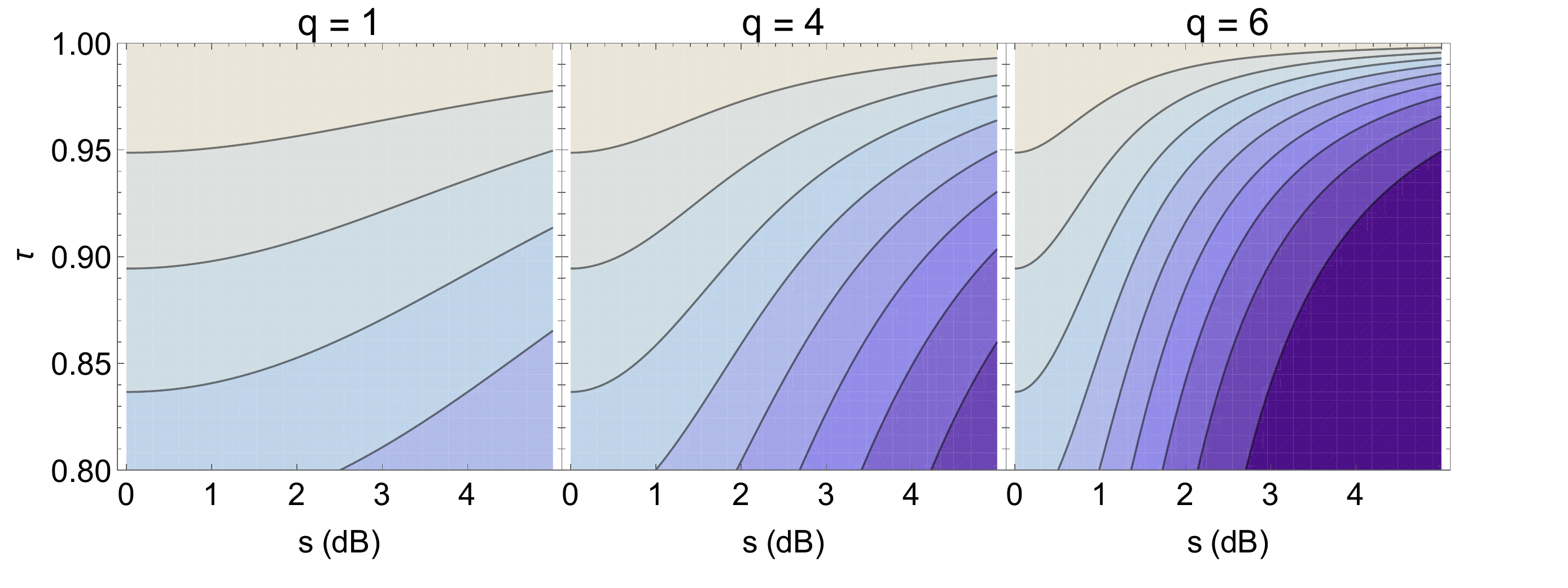}
\end{minipage}
\begin{minipage}{0.07\textwidth}\vspace{-1cm} \hspace{-1cm}
        \includegraphics[width=\textwidth]{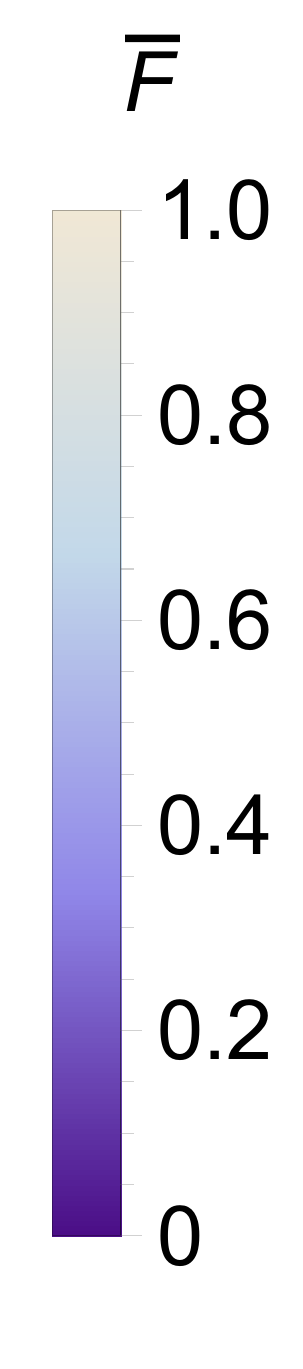} 
\end{minipage}
    \caption{\label{fig:fideContour} (Color online) Contour plots of the fidelity $\bar{\mathcal{F}}_j$ [see Eq.~(\ref{eq:FjjLoss})] for photon-subtracted states with $2^q$ modes, namely, encoding $q$ qubits, as a function of the initial squeezing and of the transmittivity of the beam splitter used to model losses.}
\end{figure*}

\begin{figure}

\begin{minipage}{0.85\columnwidth}
        \includegraphics[width=0.85\textwidth]{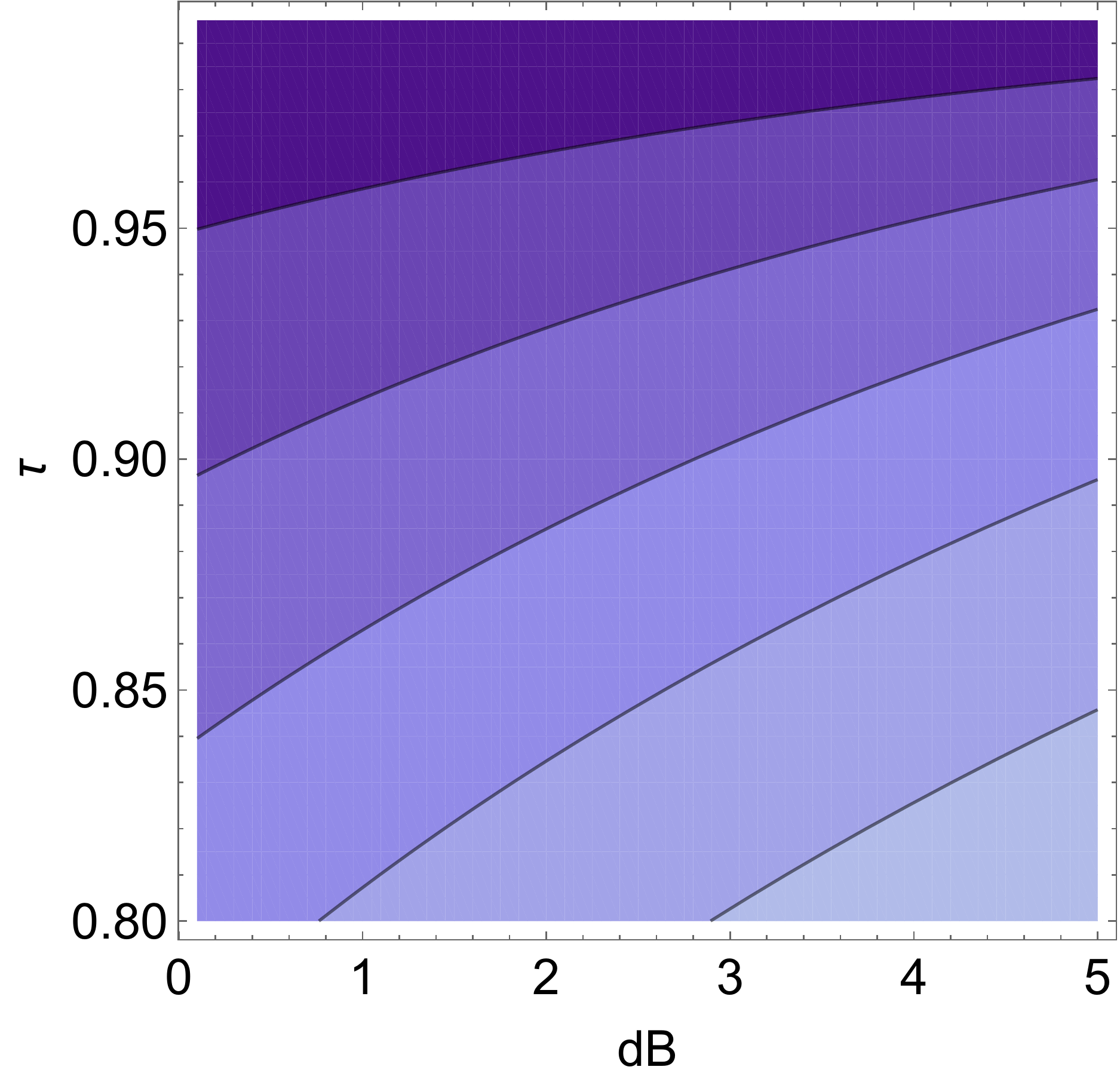}
\end{minipage}
\begin{minipage}{0.13\columnwidth}\vspace{-1cm} \hspace{-1cm}
        \includegraphics[width=\textwidth]{fideLegComb.pdf} 
\end{minipage}
    \caption{\label{fig:fideContourDiff} (Color online) Contour plots of the fidelity $\bar{\mathcal{F}}\lr{j,k}$ between two photon-subtracted states $\bar{\rho}_j$ and $\bar{\rho}_k$ after losses as a function of the initial squeezing and of the transmittivity of the beam splitter used to model losses.}
\end{figure}

Each term in this product is an overlap between a pure and a mixed state and is thus easily computed as an overlap integral between the respective Wigner functions~\cite{leonhardt1997measuring}. The results are shown in Fig.~\ref{fig:fideContour} as a function of the squeezing parameter, assumed to be the same across all modes for simplicity. Since losses degrade the squeezing, the fidelity decreases when the squeezing is increased. The effect is more severe when more modes are used to define the code space. This is also corroborated by the fidelity $\bar{\mathcal{F}}\lr{j,k} $ between two states after losses, which shows that a smaller amount of losses is sufficient to make two states less distinghishable if the initial squeezing is higher (see Fig.~\ref{fig:fideContourDiff}). In this case, since both states are mixed, we need to use the Uhlmann formula for the fidelity~\cite{wilde} and we used a truncated representation on the Fock basis for a numerical computation. 

\begin{figure}

\begin{minipage}{0.85\columnwidth}
        \includegraphics[width=0.85\textwidth]{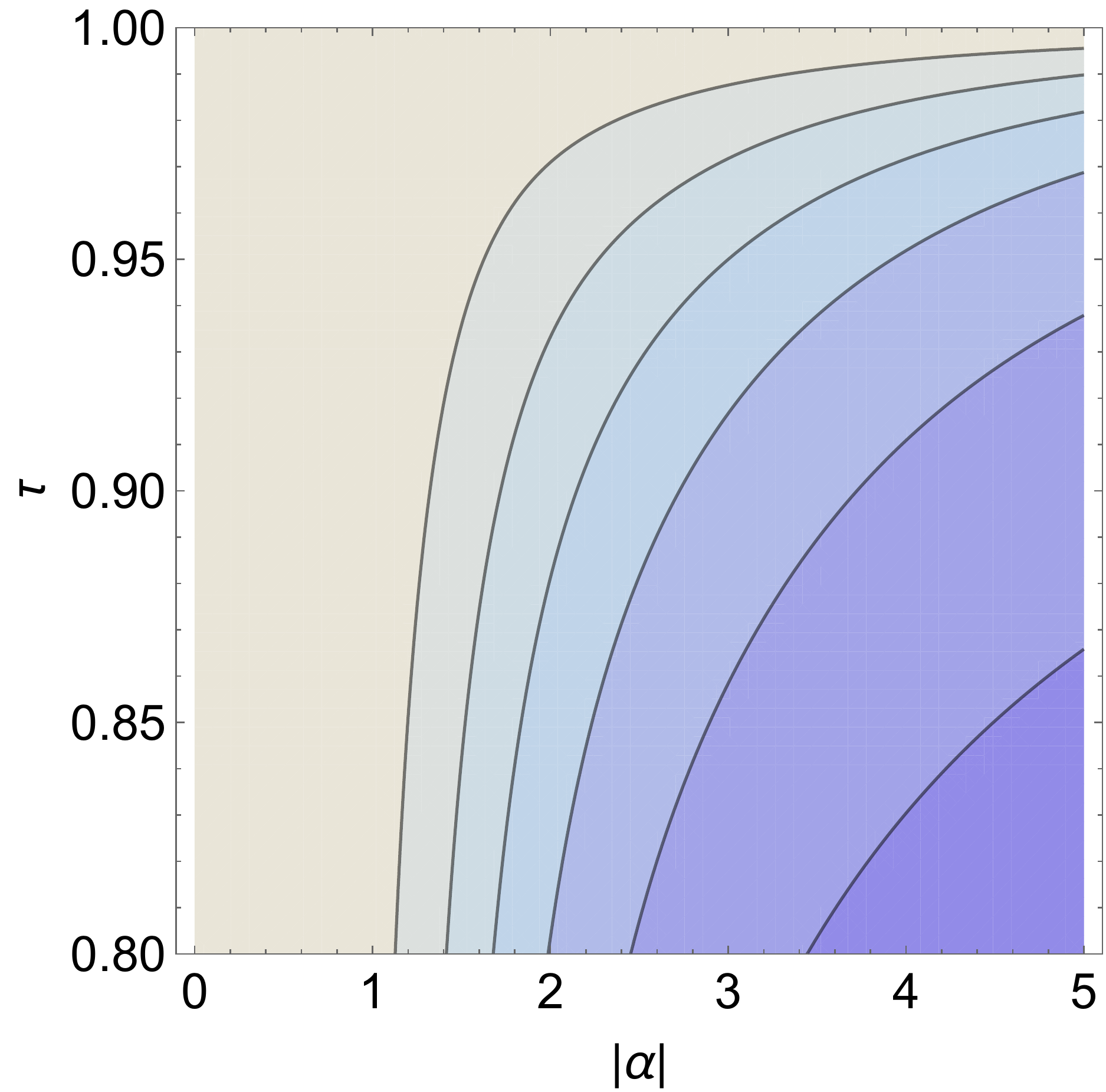}
\end{minipage}
\begin{minipage}{0.13\columnwidth}\vspace{-1cm} \hspace{-1cm}
        \includegraphics[width=\textwidth]{fideLegComb.pdf} 
\end{minipage}
    \caption{\label{fig:catLoss} (Color online) Contour plots of the fidelity between an ideal cat state with the degraded version after losses have occurred, as a function of $|\alpha|$ and of the transmittivity of the beam splitter used to model losses.}
\end{figure}

We also compare the degradation under losses of our photon-subtracted encoded states with the so called even cat state~\cite{leonhardt1997measuring}, which is a superposition of two coherent states 
\begin{equation}
\Ket{\mathrm{cat}, \alpha} \propto \Ket{\alpha} + \Ket{-\alpha}\;,
\end{equation} 
since its use has also been proposed for encoding a qubit in a single mode of radiation in several quantum-information protocols in hybrid CV-DV schemes. The results are shown in Fig.~\ref{fig:catLoss} and they have to be compared  with the first plot in Fig. ~\ref{fig:fideContour}, where a single qubit is encoded in two modes via photon subtraction. Note that since $\mathcal{L}_{\tau,1}$ is invariant under rotations of phase-space, we only need to study the impact of losses as a function of $|\alpha|$.
Although the  horizontal axes of the two pictures are not directly comparable, it is clear that the 
degradation of cat states when their amplitude increases is larger than the degradation of photon-subtracted states when the initial squeezing is increased. Moreover, while a relatively large amplitude ($|\alpha |\gtrsim 2$) of the cat states is required for $\Ket{\alpha}$ and $\Ket{-\alpha}$ to be orthogonal --- therefore for the encoding to work --- large squeezing is not required for multidimentional encoding in photon-subtracted states. Only non-zero squeezing is in fact required, in order to have non-vanishing probability of subtracting a photon from the superposition of modes. 

\section{Encoding classical data and Measuring vector distances \label{sec:classStrings}}
The encoding of classical information in a quantum state is the starting point of any quantum protocol that aims at showing any kind of advantage when compared with a classical counterpart. A key feature of the described protocol is that the encoded states can be generated simply via tuning the field gate. In particular, the encoding of a stream of classical data in a quantum state is here conditioned on one single non-deterministic photon detection, while the dimension of the involved Hilbert space is determined by the number of modes that can be deterministically squeezed in the initial MMSS.

In the following we investigate the possible advantage of computing vector overlaps (as scalar product) and vector distances via the proposed encoding. 

\subsection{Encoding two data vectors and measuring their scalar product}
Given two complex vectors $\vec{y}=\lbrace y_1,..,y_N\rbrace$ and $\vec{z}=\lbrace z_1,..,z_N\rbrace$ their components can be can be encoded in the coefficients $\gamma_j$ in Eq.~(\ref{eq:pureQudit}) via two photon-subtraction experiments that produce the two states:
\begin{equation}
\Ket[1]{PS}= \sum_j y_{j}\Ket[m_j]{s_j ^p} \bigotimes _{i\neq j } \Ket[m_i]{s_i} \equiv \sum_j y_{j} \Ket{j}\;, \label{eq:Qudit1}
\end{equation} 
\begin{equation}
\Ket[2]{PS}  = \sum_j z_{j} \Ket[m_j]{s_j ^p} \bigotimes _{i\neq j } \Ket[m_i]{s_i} \equiv \sum_j z_{j} \Ket{j}. \label{Qudit2}
\end{equation} 
If we identify with $\Pi_k$ the parity operator for the mode $k$  and with $\Ket{\Gamma}=\Ket[1]{PS}\otimes \Ket[2]{PS}$  the total state involving the two photon subtraction experiments, the probability of getting simultanoeusly the value $-1$ when the parity is measured in the same mode in the two  experiments is given by  
\begin{equation}
\mid  \Bra[1]{N_k} \Bra[2]{N_k} \Ket{\Gamma} \mid ^{2}=\mid y_{k} z_{k} \mid^{2}\;,
\end{equation} 
where $\Pi_k \Ket{N_k} =-\Ket{N_k}$.
Thus the measurement of occurrence of negative-parity coincidence in couple of modes with same index in the two experiments  estimates all the terms $\mid y_{k} z_{k} \mid^{2}$ appearing in the scalar product between the two vectors $\vec{y}$ and $\vec{z}$.  It has to be noted that the method scales linearly with the number of modes as it requires $2N$ parity measurements, while the full tomography of the total state  $\Ket{\Gamma}$ would require $2N^2$ measurements.

\subsection{Encoding the distance of two data vectors}

A second even more convenient scenario is when the difference between the two vectors is  already given as a string of classical data. We will now show that, in this case, the distance 
\begin{equation}
d_2\lr{\vec{y},\vec{z} } =\sum _i \left| y_i - z_i \right|^2 \label{eq:d2}
\end{equation} can be obtained as the variance of the quadrature operator of solely one mode, which in turn can be measured efficiently via homodyne detection. 

Before addressing the calculation of the distance between two arbitrary vectors, let us first consider $M$ independently squeezed modes and suppose we can encode a generic $M$-component vector $\vec{h}$ in the photon subtracted state: 
\begin{equation} \label{eq:psiphi}
\Ket{\psi} =  \sum _{j = 1} ^M h_j \Ket[e_j]{s_j ^p} \bigotimes _{i \neq j} \Ket[e_i]{s_i } .
\end{equation} 
It is easy to compute that \begin{equation}
\Bra{\psi} q_r \Ket{\psi} =0
\end{equation} 
for any $r$. Let us turn to the variance of the quadratures. It is clear that the computation is the same for any mode, and the choice of position or momentum is irrelevant since we did not specify the sign of the squeezing parameter. So we can consider $q_1$ without loss of generality. We define for simplicity $A \equiv \Bra{\psi} q_1 ^2 \Ket{\psi}$. We have \begin{equation} A = \sum_{j,r = 1} ^M h_j h_r ^* \lr{ \Bra[e_j]{s_j ^p} \otimes \Bra[\bar{j}]{s} } q_1 ^2 \lr{  \Ket[e_r]{s_r ^p} \otimes \Ket[\bar{r}]{s} } \label{eq:quadToMeasRaw}
\end{equation} where \begin{equation}
\Ket[\bar{j}]{s} = \bigotimes _{t \neq j} \Ket[e_t]{s_t }.
\end{equation} The terms with $j \neq r$ are zero, since each is proportional to a product of the form \begin{equation}
\Bra[e_r]{s_r ^p}\Ket[e_r]{s_r} = 0.
\end{equation} For $j=r$ there are two possibilities:  \begin{align}
j = r = 1  \implies & \Bra[e_1]{s_1 ^p} q_1 ^2 \Ket[e_1]{s_1 ^p} = 3e^{2s_1} /2 \label{eq:qExpPSub}\\ 
j = r \neq 1  \implies & \Bra[e_1]{s_1 }  q_1 ^ 2 \Ket[e_1]{s_1} = e^{2s_1} /2 .
\end{align} It follows that \begin{equation} \Bra{\psi} q_1 ^2 \Ket{\psi} = \frac{3}{2} e^{2s_1} \left| h_1\right|^2  + \frac{e^{2s_1}}{2}  \sum_{j = 2} ^M \left|h_j\right| ^2 . \label{eq:quadToMeas}
\end{equation}
The normalization of $\Ket{\psi}$ and the orthogonality of the states $\Ket{j}$ imply \begin{equation}
\sum _{j=1} ^M \left| h_j\right|^2 =1. 
\end{equation} To compute the norm of a vector of arbitrary complex numbers $\vec{x}$, one could choose a gate field corresponding to 
\begin{equation}\vec{h} = \mathcal{N}_x\lr{ \begin{array}{c} 
\beta \\ \vec{x}
\end{array} } \label{eq:betaEnc}\end{equation} where $\beta$ is a complex number chosen by the experimenter and \begin{equation}
\mathcal{N}_x = \frac{1}{\sqrt{\left|\beta\right|^2+\left|\vec{x}\right|^2}}. \label{eq:normaliz}
\end{equation}\\
The input vector $\vec{x}$ does not need to be normalized and $\beta$ can be chosen arbitrarily (as long as it is not zero, because otherwise the measured quantity no longer depends on $\vec{x}$, see Eq.~(\ref{eq:vecVar})). The normalization constant need not be computed. In a sense, it is precisely the fact that the physicality of the state enforces normalization that allows to avoid computing the vector distance explicitly, replacing its computation by a measurement. Plugging Eqs.~(\ref{eq:betaEnc}) and (\ref{eq:normaliz}) into Eq.~(\ref{eq:quadToMeas}) we find \begin{equation} \begin{aligned}
A &= \frac{3}{2} e^{2s_1} \frac{\left|\beta\right|^2}{\left|\beta\right|^2+\left|\vec{x}\right|^2}  + \frac{e^{2s_1}}{2}  \frac{\left|\vec{x}\right|^2}{\left|\beta\right|^2+\left|\vec{x}\right|^2}\\
&= e^{2s_1}\lr{\frac{\left|\beta\right|^2}{\left|\beta\right|^2+\left|\vec{x}\right|^2} + \frac{1}{2} }. \label{eq:vecVar} \end{aligned}
\end{equation} For $\left|\vec{x}\right|^2 \to 0$, the probability of subtracting from the first mode goes to $1$, so the variance of $q_1$ tends to that of a photon-subtracted squeezed mode. For $\left|\vec{x}\right|^2 \to \infty$, the probability of subtracting a photon from the first mode goes to zero, and the variance of $q_1$ tends to that of the squeezed vacuum. We find \begin{equation}
\left|\vec{x}\right|^2 = \frac{3 - 2 e^{-2 s_1} A}{2 e^{-2 s_1} A -1} \left|\beta\right|^2
\end{equation}
Note that only one quadrature has to be measured regardless of the length of the vector $\vec{x}$. This means that the norm of the vector $\vec{x}$ could be computed with a constant number of operations, namely this algorithm has $O\lr{1}$ complexity, compared to the standard $O\lr{n}$, linear in the length of the vector. It is interesting to note that the sensitivity of the measured quantity $A$ with respect to $\left|\vec{x}\right|^2$ increases with the squeezing. In fact \begin{equation}
\frac{\ddd A}{\ddd \left|\vec{x}\right|^2 } \propto e^{2 s_1}.
\end{equation} Notice that this does not depend on the squeezing of the remaining modes. Of course, said squeezing does have an impact on the overall process, as it affects the subtraction probability. This is also roughly proportional to the square of the power of the gate field, which is in turn related to $\left|\beta\right|^2 + \left|\vec{x}\right|^2$ \footnote{Encoding vectors of larger norm would require more power, so the maximum available power will ultimately limit the class of vectors that can be encoded in this scheme. Also, if the power of the gate is too large, the probability of more than one subtraction becomes non-negligible.}. The expected waiting time to have enough subtraction events to collect a reasonable statistics for an estimate of $A$ will then be longer for small values of $\left|\vec{x}\right|^2$ but it is anyway bounded because of the finite value of $\beta$. 

To evaluate the distance between two vectors $\vec{y}$ and $\vec{z}$, one could use the full $N = m+1$ modes system to encode \begin{equation}\vec{h} = \lr{ \begin{array}{c} 
\beta \\ \vec{y} - \vec{z}
\end{array} } \end{equation}(where $m$ is the length of $\vec{y}$ and $\vec{z}$) and then measure the variance of $q_1$. Note that $N = m+1$ squeezed modes are needed, but this has to be compared with the linear scaling one would have in the dimension of the classical registers needed to store the vectors in classical algorithms and is a space complexity problem rather than time.

As for the encoding procedure, if the vectors $\vec{y}$ and $\vec{z}$ are given in the form of frequency-shaped strong coherent pulses, the gate pulse for the subtraction needed to encode the coefficients $y_i - z_i$ on the photon-subtracted MMSS can be obtained as follows. First apply to each a frequency-dependent attenuation accounting for the subtraction probability from each of the squeezed modes. This is known beforehand from the characterization of the MMSS and the photon subtractor and can be done with a fixed device (for example a pulse shaper). Then mix the two beams on a balanced beam splitter. The output mode of the beam splitter containing the difference of the amplitudes of the input beams is then used as a gate. The complexity of this procedure is again independent of number of modes.

\section{Conclusions \label{sec:end}}

The generation of photon-subtracted optical states has been the subject of intensive experimental efforts for more than a decade \cite{Ourjoumtsev2006a, NeergaardNielsen2006, Ourjoumtsev2006b, Ourjoumtsev2007a, Zav08, Allevi2010, Yukawa2013, Dong2014, Sychev2017}. This is motivated by both applicative and fundamental considerations, given that the access to these type of states could improve the performances of a variety of quantum information tasks \cite{kim2008} --- including estimation \cite{Adesso2009} and teleportation protocols \cite{Opatrny2000, Cochrane2002, Olivares2003} --- and allow for homodyne-based loop-hole free non-locality tests \cite{Nha2004, Garcia-Patron2004, ferraro2005nonlocality, acin2009tests, Etesse2014a}. More in general, photon-subtracted states can be used as resources for tasks that require quantum non-Gaussianity or Wigner negativity \cite{Takagi2018, albarelli2018resource}, and in fact they have been proposed as building blocks to implement universal non-Gaussian operations \cite{Arzani2017} and hard-to-sample non-universal dynamics \cite{Olson2015}. In addition, arbitrary single-mode quantum states can be engineered when multiple feed-back controlled photon-subtraction operations are applied sequentially \cite{Fiurasek2005}. Here we have shown how, in a multi-mode setting, they could be used for general quantum encoding purposes. Specifically, we introduced a multidimensional quantum encoding which is based on multimode CV states of light. The code dimension is determined by the number of light modes which can be simultaneously occupied by 
squeezed vacua, and the information is encoded on the coefficients of the superposition of photon-subtraction events, which are in turn coherently applied to the squeezed modes and triggered by a single photon event.

The encoding is a generalization of the multirail encoding and it coincides with the latter in the limit of zero squeezing. A noteworthy difference of our scheme with respect to the  standard multirail approach is that it requires only one single photon detector (at the encoding stage), as we propose the use of homodyne detection.  Moreover, the adjustment of the encoding coefficients requires only the control on the spectral components of a gate beam in a coherent state (ultimately controlled via a spatial light modulator) and not the arrangement of several interferometric parameters via optical components (beam splitters, phase shifters). Notice that the encoding does not require a large amount of squeezing per mode: any non-zero squeezing is sufficient to ensure a measurable rate of subtraction events --- which is in any case mainly driven by the amplitude of the gate field. Our results suggest that large values of squeezing are in fact detrimental for the encoding, making the computational basis states less robust to losses. This implies a trade-off between resilience to losses and the rate of generation, as the number of subtraction events in a given time is proportional to the mean photon number in the subtraction mode, the power of the gate field and the square of the nonlinear susceptibility \citep{Averchenko14}. Low squeezing values imply low photon numbers and thus lower rates. This can to some extent be compensated, for fixed nonlinearity, increasing the power in the pump beam. However, at high gating power spurious effects may appear, such as dark counts on the heralding detectors due to second-harmonic generation of the gate field, that degrade the quality of the subtraction process.

We stress that our investigations are mostly motivated by the experimental readiness of much of the technology required. The coherent single photon subtraction has been demonstrated on a space of 16 modes  \cite{Ra17},  and applied on multimode squeezed and entangled states \cite{inprep}. The production of multimode quantum states involving a large number modes, up to $60$ for frequency modes and up to $10^6$ for temporal modes has already been demonstrated in different experimental setups  \cite{Roslund14,Chen14,million}. The tailoring of nonlinear processes  based on  spectral and temporal modes represents an active area of research on its own, as reviewed in \cite{Ansari18},  and we can reasonably expect the extension of coherent single-photon subtraction to a larger number of squeezed modes. Besides the number of non-vacuum modes, the dimension of the code-space is only limited by the specifics of each setup (e.g. the number of up-conversion modes that can be phase-matched simulataneously and the resolution of the pulse-shaping device used to produce the gate pulse). It is thus interesting to explore new ways to exploit the resulting states for quantum information processing. The results we report suggest that besides producing interesting high-dimensional quantum states (qudits), of which we also give some explicit examples in Appendix~\ref{App:states}, the tunability of the subtraction mode can be exploited to map arbitrary classical data to quantum states, as well as process them. We showed this by providing in Sec.~\ref{sec:classStrings} two schemes to compute the distances of two encoded vectors, which may have a wide range of applications, for example in clustering algorithms for machine learning. The coefficients can be transferred to the multimode state of the field that shapes the subtraction mode, and the norm or the distance are shown to be proportional to the mean values of simple observables: either the parity or the square of a quadrature operator. 

In summary, these  findings represent a first step in the use of multimode photon-subtracted squeezed states for quantum information processing, with the potential to lead to more advanced tasks, such as universal quantum computation. We only briefly consider this last application in Appendix~\ref{App:gates}. Although we can formally construct a universal set of gates, their experimental realization lies outside the reach of current experimental capabilities. The possibility to construct more experimentally accessible sets of transformations remains an open question that we leave to further investigations.

\section*{Acknowledgments}

We thank Young-Sik Ra for giving us access to the experimental data we used to compute the fidelity $\tilde{\mathcal{F}}\lr{j,k}$ in Sec.~\ref{sec:fidePhotoSub}. The authors are grateful to Nicolas Treps for interesting discussions. FA acknowledges financial support from the French National Research Agency project ANR-17-
CE24-0035 VanQuTe. AF acknowledges support from EPSRC (grant EP/P00282X/1). 

\appendix

\section{Examples of encoded states}
\label{App:states}

We can easily see that interesting classes of states can be obtained by simply tuning the coefficients $\gamma_j$ in Eq.~(\ref{eq:psub_MMSS}), which are controlled by the classical gating field. To fix the ideas, consider $2^n$ independently squeezed modes and the corresponding states $\Ket{j}$ in which a photon has be subtracted from mode $e_j$ (see Eq.~{\ref{eq:psub_basis}). These could represent an $n$-qubit system in which $\Ket{j}$ corresponds to the binary representation of $j$. For example for $n=2$, neglecting normalization after photon subtraction \begin{align} \begin{split}
\Ket{j = 0} =& \sigma_0  \Ket[m_0]{s_0} \Ket[m_1]{s_1} \Ket[m_2]{s_2}\Ket[m_3]{s_3}\leftrightarrow \Ket{00} \;,\\
\Ket{j = 1} =&  \Ket[m_0]{s_0}  \sigma_1  \Ket[m_1]{s_1} \Ket[m_2]{s_2}\Ket[m_3]{s_3}\leftrightarrow \Ket{01} \;,\\
\Ket{j = 2} =&  \Ket[m_0]{s_0} \Ket[m_1]{s_1} \sigma_2  \Ket[m_2]{s_2}\Ket[m_3]{s_3}\leftrightarrow \Ket{10} \;,\\
\Ket{j = 3} =&  \Ket[m_0]{s_0} \Ket[m_1]{s_1} \Ket[m_2]{s_2} \sigma_3 \Ket[m_3]{s_3}\leftrightarrow \Ket{11} \;.\end{split}
\end{align} 
A two-qubit encoded cluster state \cite{oneWayQC}, for example, would then correspond to 
\begin{equation} 
\begin{aligned}
\Ket{G_2} &= \frac{1}{2} \lr{\Ket{00} + \Ket{01} + \Ket{10} - \Ket{11}} \\ &= \frac{1}{2} \sum _j \lr{-1}^{\delta_{j3}} \Ket[e_j]{s_j^{\lr{p}}} \bigotimes_{i\neq j} \Ket[e_i]{s_i}
\end{aligned} 
\end{equation} 
where $\delta_{jk}$ denotes the Kronecker delta. Cluster states have plenty of applications in quantum information, for example they are known to be universal resources for measurement-based quantum computing if combined with an appropriate set of measurements. These usually include so-called Pauli measurements and at least one non-Clifford gate \cite{Gottesman1998}. The latter requires measurements that are usually considered harder to implement. This difficulty can be circumvented using magic state injection \cite{magicState}, which roughly involves coupling an eigenstate of the desired non-Clifford operator to the cluster state. It is easy to imagine that the resulting extended resource state can again be written as a MMPS state by tuning the coefficients in Eq.~(\ref{eq:psub_MMSS}). The catch is of course that it is difficult to write, let alone realize, even Pauli measurements on the encoded states. Nevertheless, the present manuscript may motivate the search for an experimental scheme to perform such measurements and, if such a scheme was found, a plethora of protocols could readily be implemented as the resource states are already available. We conclude by giving two more examples of interesting states that could be produced with the present setup. Hypergraph states are a generalization of cluster states containing edges between more than two qubits that were introduced to study the entanglement properties of some quantum algorithms \cite{hypergraph}. An edge between three qubits may be understood as the result of applying a $Z$ on the third if the state of both the other two is $| 1 \rangle$. The smallest non-trivial hypergraph involves three qubits and the encoded version can be written \begin{equation}
\left| \mathrm{Hyp}3\right> = \frac{1}{\sqrt{8}} \sum _{j=0}^{7} \lr{-1}^{\delta_{j7}} \Ket[e_j]{s_j^{\lr{p}}} \bigotimes_{i\neq j} \Ket[e_i]{s_i}.
\end{equation} Finally, fingerprinting designs a class of communication protocols where two parties, Alice and Bob, have two strings of $n$ bits $a$ and $b$ and a third party, Charles, has to decide whether $a=b$. Charles can communicate with Alice and Bob, but Alice and Bob cannot communicate. Their goal is to send the minimum amount of information to Charles still allowing him to decide whether $a=b$ with small error probability. Quantum mechanics allows an exponential reduction \cite{qFingerprint} of the amount of information that Alice and Bob have to send by using an error correcting code $\mathcal{E}:\left\{0,1 \right\}^n \to \left\{0,1 \right\}^M$ such that $\mathcal{E}\lr{a} = x$, $\mathcal{E}\lr{b} = y$ and then encoding $x$ and $y$ in the states \begin{align}
\Ket[]{h_x} &= \frac{1}{\sqrt{M}} \sum _{j=0}^{M} \lr{-1}^{x_j} \Ket[]{j} \\
\Ket[]{h_y} &= \frac{1}{\sqrt{M}} \sum _{j=0}^{M} \lr{-1}^{y_j} \Ket[]{j}
\end{align} where $x_j$ ($y_j$) is the value of the $j$th bit of $x$ ($y$). This also nicely matches our representation of a MMPS states.

\section{Universal set of gates}
\label{App:gates}

 We have seen that a single photon-subtraction acting coherently on $M$ modes can initialize an $M$-level system to an arbitrary state. Let us now briefly address the question of the operations required to perform arbitrary quantum processing of the information stored on $M$ modes via the encoding here introduced. For the sake of this Appendix, we will only focus on the case of $M=2^d$ (namely, on a high dimensional system that emulates a multi-qubit one). 

At the logical level, a universal set of qubit gates consist of all possible single-qubit unitaries plus an entangling two-qubit gate. A possible  finite universal set of gates is given by $\{H, T, CZ \}$: it is composed of the Hadamard gate, the $T$ gate (a rotation of angle $\frac\pi4$ around the Pauli $Z $ axis), and the control-phase $CZ$ respectively \cite{NChuang}. We will only introduce here the definitions of such gates in terms of the encoding, leaving the analysis of their possible implementation to future studies.

Considering the definition of the parity operator given in Eq.~\eqref{eq:parity_tilde}, the $T$ gate acting on a logical qubit composed of two modes is expressed as
\begin{equation}
T=\tilde{\Pi}_2^{\pi/4}\;.
\end{equation}
The Hadamard gate, again acting on a qubit encoded on two modes, corresponds to the following transformation:
\begin{align}
\Ket{s^p_1,s_2} &\rightarrow \frac{1}{\sqrt2}
\left(
\Ket{s^p_1,s_2}+\Ket{s_1,s_2^p}
\right)
 \;, \\
\Ket{s_1,s_2^p} &\rightarrow \frac{1}{\sqrt2}
\left(
\Ket{s^p_1,s_2}-\Ket{s_1,s_2^p}
\right)\;.
\end{align}

Finally, the $CZ$ between two logical qubits encoded on four modes can be written as:
\begin{equation}
CZ= \sum_{j=0}^1\vert j  \rangle  \langle j \vert   \otimes \Big(\bigotimes _{k=0} ^{1} \Pi_k^{k j}\Big)\;.
\end{equation}
 

\bibliography{./pSubQdit-arxiv-postReview}{}

\end{document}